\documentclass[11pt]{article}

\usepackage[margin=1in]{geometry}

\usepackage{graphicx}
\usepackage[authoryear,round]{natbib}
\usepackage{authblk}
\usepackage{mathtools}
\usepackage{etoolbox}
\usepackage[section]{placeins}

\usepackage{framed,multirow}
\usepackage{latexsym}
\usepackage{amsthm,amsmath,amssymb,amsfonts,amscd,xspace,pifont,bm}
\usepackage{url}
\usepackage{xcolor}
\usepackage{tabularray}
\usepackage[short]{optidef}
\usepackage{gensymb}
\usepackage[]{footmisc}
\usepackage{siunitx}
\usepackage[colorlinks=true,linkcolor=blue,citecolor=blue,urlcolor=blue]{hyperref}

\newcommand{\ze}{Z\kern-0.45emZ}
\newcommand{\esp}{I\kern-0.37emE}
\newcommand{\N}{I\kern-0.37emN}
\newcommand{\one}{ {\rm 1\kern-0.19eml} }
\newcommand{\realset}{I\kern-0.37emR}

\newcommand{\bx}{\boldsymbol{x}}

\newcommand{\bX}{\boldsymbol{X}}
\newcommand{\bB}{\boldsymbol{B}}
\newcommand{\bC}{\boldsymbol{C}}

\newcommand{\bb}{\boldsymbol{b}}

\newcommand{\bq}{\boldsymbol{q}}
\newcommand{\bd}{\boldsymbol{d}}
\newcommand{\bc}{\boldsymbol{c}}

\makeatletter
\newcommand{\ssymbol}[1]{^{\@fnsymbol{#1}}}
\makeatother

\DeclarePairedDelimiterX\Card[1]\lvert\rvert{%
  \ifblank{#1}{{-}}{#1}%
}
\DeclarePairedDelimiter{\norm1}{\lVert}{\rVert}

\def\red{\textcolor{black}}
\def\blue{\textcolor{black}}
\definecolor{newcolor}{rgb}{.8,.349,.1}

\title{Robust brain MRI image classification with SIBOW-SVM}

\author[1]{Liyun Zeng}
\author[2]{Hao Helen Zhang\thanks{Corresponding author. Email: \texttt{haozhang@arizona.edu}}}

\affil[1]{Statistics and Data Science GIDP, University of Arizona, Tucson, Arizona 85721, USA}
\affil[2]{Department of Mathematics, University of Arizona, Tucson, Arizona 85721, USA}

\date{} 

\begin{document}

\maketitle

\begin{abstract}
Primary Central Nervous System tumors in the brain are among the most aggressive diseases affecting humans. Early detection and classification of brain tumor types, whether benign or malignant, glial or non-glial, is critical for cancer prevention and treatment, ultimately improving human life expectancy. Magnetic Resonance Imaging (MRI) is the most effective technique for brain tumor detection, generating comprehensive brain scans. However, human examination can be error-prone and inefficient due to the complexity, size, and location variability of brain tumors. Recently, automated classification techniques using machine learning methods, such as Convolutional Neural Networks (CNNs), have demonstrated significantly higher accuracy than manual screening. However, deep learning-based image classification methods, including CNNs, face challenges in estimating class probabilities without proper model calibration \citep{guo2017,minderer_revisiting_2021}. In this paper, we propose a novel brain tumor image classification method called SIBOW-SVM, which integrates the Bag-of-Features model with SIFT feature extraction and weighted Support Vector Machines. This new approach can effectively extract hidden image features, enabling differentiation of various tumor types, provide accurate label predictions, and estimate probabilities of images belonging to each class, offering high-confidence classification decisions. We have also developed scalable and parallelable algorithms to facilitate the practical implementation of SIBOW-SVM for massive image datasets. To benchmark our method, we apply SIBOW-SVM to a public dataset of brain tumor MRI images containing four classes: glioma, meningioma, pituitary, and normal. Our results demonstrate that the new method outperforms state-of-the-art techniques, including CNNs, in terms of uncertainty quantification, classification accuracy, computational efficiency, and data robustness.
\end{abstract}


\medskip

\newpage
\section{Introduction}
\label{sec:introduction}
Brain tumors are abnormal cell growths within the human brain, classified as either malignant (cancerous) or benign (noncancerous). The three major types are gliomas, meningiomas, and pituitary tumors. Gliomas, typically  malignant, fast-growing, and invasive, originate from glial cells (astrocytes, oligodendrocytes, and ependymal cells)
that support neurons. These tumors can significantly impact brain functions and pose life-threatening risks \citep{deangelis_brain_2001}. In contrast, meningioma 
and pituitary tumors are usually benign and slow-growing.  
Meningiomas arises from meninges, membranes surrounding the brain and spinal cord, while pituitary tumors develop in the pituitary gland behind the nasal cavity \citep{rg_2017170037}. Although these tumors generally do not spread to other body parts, they can affect nearby brain tissues, nerves, or blood vessels, potentially causing severe disabilities. As brain tumors grow, they can exert pressure on surrounding nerves, affecting brain functions and overall patient health. Early detection and accurate classification are crucial for effective treatment and improved survival rates \citep{hishii_diagnosis_2019}. However, brain tumor screening is challenging even for experienced radiologists due to the physiological complexity of tumors \citep{Afshar2019CapsuleNF}.

Magnetic Resonance Imaging (MRI) is a primary diagnostic tool for brain tumors \citep{amin_distinctive_2020} \blue{and guides neuro-surgeons in precise tumor removal \citep{asthagiri2007advances}. This painless procedure uses powerful magnets, radio waves, and computer technology to produce detailed images of brain structures, including blood vessels, skull, and facial bones. In clinical practice, MRI scans reveal various brain abnormalities, such as inflammation and swelling, structural issues, abnormal growths or masses (brain tumors), fluid leaks, hemorrhage, and white matter disease. Neurosurgeons use MRI for diagnosing new neurological conditions, monitoring existing ones, and surgical planning. Intraoperative MRI (iMRI) \citep{foroglou2009intra} enhances surgical precision, allowing real-time guidance during tumor removal and treatment of conditions like epilepsy \citep{duncan2016brain}.} 

Unlike tumors in other body parts, brain tumor surgeries and biopsies are complex and risky procedures. Thus, reliable MRI-based classification is crucial for early diagnosis. \blue{Computer-aided diagnosis systems assist radiologists in rumor detection, but challenges persist and caused by issues like morphological similarities between different tumor types \citep{histed2012review}, difficulty in visualizing small or deep-seated tumors \citep{iijima2015microrecording}, tumor heterogeneity \citep{keane2021multifaceted}, image directions, and surrounding tissue noise \citep{philip2022artificial}. For example, an aggressive glioblastoma may resemble a metastatic tumor, brain abscess (infection), or even multiple sclerosis \citep{mandel2016primary}. In summary, while MRI technology significantly advances brain tumor diagnosis and treatment, accurate identification and classification remain challenging. Motivated by this, our research aims to improve imaging techniques and computer-aided diagnosis systems for more effective brain tumor detection and treatment. } 

In recent years, advancements in computational technology and machine learning methods have significantly improved MRI-based tumor screening \citep{currie_machine_2019}. Various machine-learning techniques designed for computer vision and image classification now provide radiologists with abundant information and objective recommendations for tumor classification. \red{Machine-aided auto-segmentation of brain MRI images can classify one slice of an image (2D), five consecutive slices of an image (2.5D), or the entire volume of an image (3D). Recent research suggests that 3D models can achieve higher classification accuracy compared to 2.5D or 2D models. However, this improved performance comes at the cost of significantly increased computational memory requirements, often more than 20 times that of their lower-dimensional counterparts. This limitation restricts the use of 3D models for large data sets in practice \citep{avesta2003}. The trade-off between accuracy and computational resources highlights the importance of model selection in classification.} 

Convolutional Neural Networks (CNNs) have gained widespread popularity for their high accuracy in classifying both \red{3D and 2D} brain tumor MRI images, while being able to extract intricate features from images \citep{wang8840843,ayadi_deep_2021}. \blue{CNNs and their variations such as VGG \citep{engproc2023059046}, AlexNet \citep{ieee10014115}, ResNet \citep{ieee8882973,GOPI2024123}, GoogLeNet \citep{ieee9501999}, DenseNet \citep{ieee9972116}, and EfficientNet \citep{BERNAL201964,ieee8697915,diagnostics12081850, babu2023detection} play a major role in recent clinical applications of brain tumor identification with MRI images.
As of \citet{litjens_survey_2017}, there were about 220 research works related to deep learning for medical images in 2016, with 190 of them adopting CNNs as their methodology. Furthermore, the implementation of parallel CNNs on the NVIDIA CUDA GPU architecture has substantially accelerated the processing of massive image data and model training, making CNNs the gold standard for image classification \citep{Sultan8723045}. In literature, CNN techniques perform well for most brain tumor classification tasks in real-world clinical diagnosis, including two-class classification, e.g. normal vs. tumorous \citep{khan2020brain, el2021brain}, higher grade glioma (HGG) vs. lower grade glioma (LGG) \citep{yang2018glioma,decuyper2021automated}, and three-class classification, e.g. normal vs. LGG vs. HGG \citep{tandel2020multiclass}, and Meningioma vs. Pituitary tumor vs. Glioma \citep{ismael2020enhanced,swati2019brain}. However, classification performance for more than three classes remains poor and requires further research \citep{kv2019glioma, xiao2021synergy,tandel2020multiclass}.} 

CNNs have demonstrated remarkable success in classifying large-scale image datasets such as ImageNet, which contains over one million images \citep{russakovsky_imagenet_2015}. However, applying deep CNNs to medical image classification, particularly with limited data, presents a few main challenges: (i) overfitting and convergence issues: training deep CNNs on small datasets often results in models that perform well on training data but fail to generalize to new, unseen examples; (ii) computational demands: hyperparameter tuning, crucial for optimizing model performance, adds substantial computational costs \citep{Tuba2021}; (iii) probability estimation: in the context of brain tumor screening, accurate probability estimation is critical for quantifying prediction uncertainty. Recent studies indicate that deep CNNs may struggle to provide precise probability estimations \citep{guo2017, kumar_trainable_2018, minderer_revisiting_2021}. These challenges underscore the need for specialized approaches when applying CNNs to medical imaging tasks, particularly with limited data availability. 

A promising alternative to CNNs for medical image classification is the class of weighted Support Vector Machines (wSVMs) \citep{wang_multiclass_2019}. This SVM-based learning approach is specially designed for probability estimation in $K$-class problems, denoted the 
number of classes is denoted as $K$. The wSVMs offer several advantages: flexibility and robustness in estimating posterior class probabilities, accuracy in probability estimation, and polynomial-time complexity in $K$. Recent works by \citet{zeng_wsvms_2022} has further enhanced wSVMs by introducing two new learning schemes: baseline learning and one-vs-All (OVA) learning, to improve computational efficiency and probability estimation accuracy. In particular, the baseline learning scheme achieves optimal linear complexity, and the OVA scheme demonstrates superior estimation accuracy in numerical studies. These wSVM approaches have shown notable performance improvements compared to traditional methods such as multinomial logistic regression, multiclass linear discriminant analysis, classification trees, Random Forest, and XGBoost.

In this article, we propose a novel approach called \textbf{SIBOW-SVM} for both image classification and probability estimation, based on weighted Support Vector Machines(wSVMs). \red{While we demonstrate the new algorithm using 2D image classification problems to highlight its low-memory cost and computational efficiency, the SIBOW-SVM framework can be extended to 3D models by aggregating consecutive image slices in three-dimensional space.} The SIBOW-SVM learning framework consists of three main steps: (i) feature extraction: using scale-invariant feature transformation (SIFT) technique \citep{lowe_distinctive_2004}, (ii) feature processing: applying the Bag-of-Feature (BoF) model \citep{Hiba2016} for codebook generation, feature encoding, and pooling, (iii)  classification and probability estimation employing wSVMs. 
The key advantages of SIBOW-SVM include high accuracy, computational efficiency and scalability, and robustness.
Its computational cost is comparable to CNNs without NVIDIA CUDA GPU acceleration. Furthermore, its divide-and-conquer nature allows for significant speed-up through parallel computing using GPUs, multi-core processors, high-performance computing (HPC) clusters, or massively parallel computing (MPP) systems. We will evaluate the performance of SIBOW-SVM on public image data sets and compare its results with other methods, including CNNs.  

The remainder of this paper is organized as follows. Section \ref{sec:systems} describes image classification systems and pre-processing techniques, and introduces the brain tumor MRI image data set. Section \ref{sec:3.1} presents the proposed SIBOW-SVM methodology, elucidating its computational algorithm and practical implementation details. Section \ref{sec:res4} presents numerical results and summarizes key findings from the real data analysis. Section \ref{sec:cond} discusses concluding remarks and future research directions. 

\section{Image Classification Systems} \label{sec:systems}
\subsection{Image Classification Systems}
Image classification involves a sequence of systematic steps, from image preparation to model building. The process can be outlined as follows. The first step is image pre-processing, which consists of image labeling and standardization. Standardizing raw images to a uniform pixel size is essential for both model training and cross-method evaluation. For fair assessment and comparison of different methodologies, we randomly split the entire dataset into two subsets: the training set $\mathbb{S}_{train}$ (80\%) and the test set $\mathbb{S}_{test}$ (20\%). 

\begin{figure}[!htbp]
\centering
     \includegraphics[trim=0.2cm 5.5cm 0.2cm 2cm, width=0.95\textwidth, height=0.7\textheight, keepaspectratio]{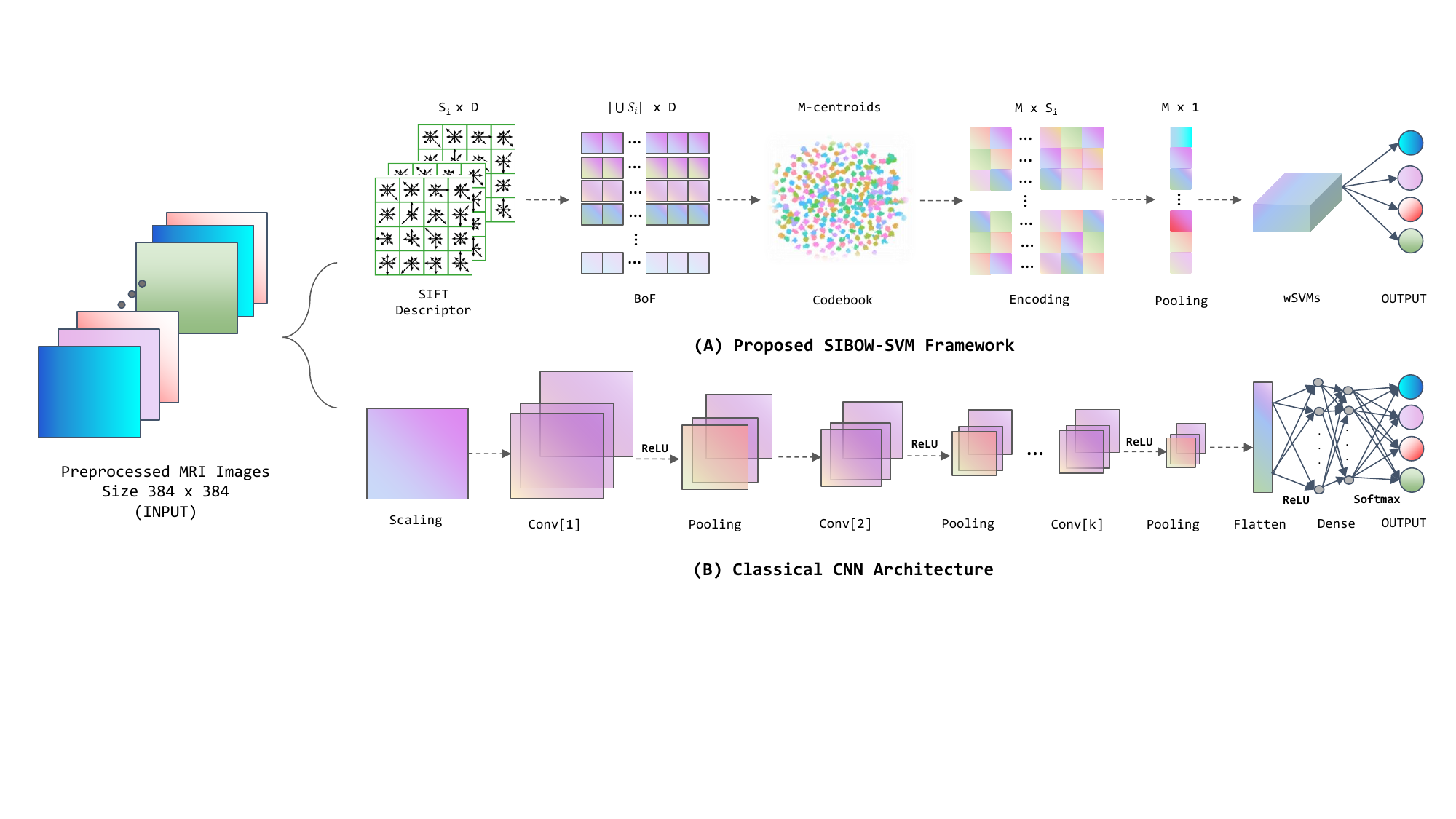}
\caption{Brain tumor MRI image classification systems for: (A) proposed SIBOW-SVM framework (upper) and (B) CNN network architecture (lower)}
\label{fig:figure_1}
\end{figure}

In Figure \ref{fig:figure_1}, we illustrate the image classification systems employed in both the CNN network architecture and our proposed SIBOW-SVM. Both systems utilize $\mathbb{S}_{train}$ for training and $\mathbb{S}_{test}$ for performance evaluation. The lower row (b) depicts the CNN network architecture, which begins with the scaling of pixel values, followed by multiple convolutional and pooling layers using the ReLU activation function. The pooled features are then flattened and fed into a fully connected dense network, which uses the softmax function for classification. The SIBOW-SVM framework is presented in the upper row (b). It starts with SIFT feature extraction from each image, and these features are combined to create a large Bag of Features (BoF) feature pool. The training images are then described through histograms of local descriptors. The subsequent step is codebook construction by clustering the BoF feature pool (e.g. using $k$-means algorithm) to generate $M$ centroids as \emph{visual words}. Each image is then encoded using this codebook, resulting in a list of \emph{visual codes} to represent local descriptors. These visual codes can be seen as local histograms of $M$ dimensions, corresponding to the size of the codebook. The final image feature is an $M$-vector obtained by pooling and normalizing the \emph{visual code} encoding, which is then used as input to train weighted Support Vector Machines (wSVMs) for classification and probability estimation \citep{zeng_wsvms_2022}.  

\subsection{Data Set and Pre-Processing}
\label{ssec:imgprepro}
We utilize a public dataset from Kaggle (\url{www.kaggle.com}), \red{available at \url{https://github.com/SartajBhuvaji/Brain-Tumor-Classification-DataSet.git}}, referenced in \citet{sartaj_kaggle_2020}. This dataset consists of 3,264 2D brain tumor MRI scans in JPEG format, categorized into four classes: Glioma (926 images), Meningioma (937 images), Pituitary (901 images), and Normal (500 images). These images are acquired across three different planes: sagittal, axial, and coronal. In our analysis, we \red{merged the original training and testing data, then} randomly split the combined dataset into a training set (2,606 images, 80\%) and a test set (658 images, 20\%). The proportion of each class was maintained in both sets. 

\begin{figure}[!htbp]
\centering
     \includegraphics[clip, trim=0.3cm 0.5cm 0.5cm 0.5cm, width=0.85\textwidth, height=0.32\textheight, keepaspectratio]{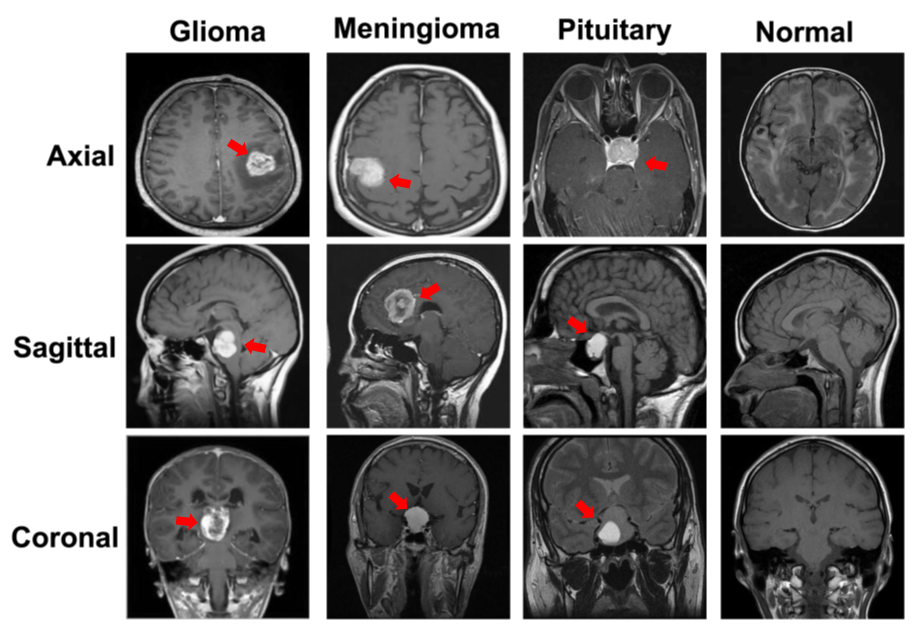}
\vspace{-6pt}
\caption{Representation of normalized brain tumor MRI images of three brain tumor types and normal brain tissues in the 
sagittal, axial, and coronal planes. Red arrows point to the tumor lesions.}
\label{fig:figure_2}
\end{figure}

The original images in this data set vary in size and resolution, ranging from $1280 \times 1280$ to $128 \times 128$ pixels, and include different shapes (squares and rectangles). In the standardization process, we normalize all images to a resolution of $384 \times 384$ pixels, ensuring a uniform square shape. For the CNN network, we scale the pixel values to the range $[0,1]$. It is important to note that no data augmentation was applied to this dataset. Fig. \ref{fig:figure_2} presents examples of the three types of brain tumors and normal brain tissues, depicted in the sagittal, axial, and coronal planes. Each image distinctly marks the tumor lesions with a red arrow.


        
\section{Methodology}
\label{sec:3.1}

\subsection{Convolutional Neural Networks (CNNs)}\label{ssec:3.1}

Convolutional Neural Networks (ConvNets or CNNs) are state-of-the-art tools for image classification, particularly effective for massive image datasets. Their advantages include scalability, high classification accuracy, and efficient optimization (e.g., ImageNet)\citep{russakovsky_imagenet_2015}. They can be further accelerated through parallel computation using NVIDIA CUDA GPUs \citep{yadav_deep_2019}. 

A CNN network model consists of two major components: feature extraction and classification. In feature extraction, successive convolutional layers and the ReLU activation function work in tandem to extract meaningful features at various dimensions, while using pooling layers to reduce the feature map size. The classification is performed by fully connected dense layers, using extracted features to estimate class probabilities and a softmax function for the final output (see Fig.\ref{fig:figure_1}). Each neuron in these layers is assigned a weight and a bias, which are learned during the training process. The input to each neuron is a weighted sum of inputs, and the output is obtained through a non-linear transformation via activation functions like ReLU. 

The convolutional layer involves multiple hyperparameters,  including (1) filter, which determines the output space dimensionality and acts as a template to find similarities in the input image;  (2) kernel size, which specifies height and width of the 2D convolution window; (3) stride, which defines convolution strides along height and width;  (4) the padding method, which is either ``valid" (no padding) or ``same" (zero padding evenly around the input); (5) activation function, typically ReLU for image classification; and (6) kernel regularizer, which applies regularization to the kernel weights matrix, such as $\ell^1$ (Lasso), $\ell^2$ (Ridge), or $\ell^1+\ell^2$ (Elastic Net). To stabilize learning in each activation layer, batch normalization can be added to normalize each mini-batch (mean close to 0, standard deviation close to 1), improve CNN performance, and accelerate training \citep{garbin_dropout_2020}. 

The pooling layer is a critical component of CNNs. It reduces computational load by summarizing features in a region of the feature map from a convolutional layer, downsizes the feature map, and reduces the number of learning parameters. Post-convolution, either max or average pooling, is employed to reduce the input along its spatial dimensions for each channel, with the pooling window shifted by strides along each dimension. The output of the pooling layer feeds into the subsequent convolutional layer, with the final pooling layer being flattened into a feature vector for classification by the fully connected neural network layer (Fig. \ref{fig:figure_1}). 

The fully connected (or dense) layer includes a flattened image feature vector as the input, multiple hidden layers with varying neuron sizes, and an output layer where the number of neurons is equal to the number of classes, (e.g., $K=4$ in our example). Each neuron is interconnected across adjacent layers. We use ReLU as the activation function between input and hidden layers, softmax for class probability estimation, and argmax to class label prediction. The key hyperparameters include the number of hidden layers (depth) and the number of neurons in each layer. A CNN network can be trained using algorithms like back-propagation and gradient descent. Commonly used optimizers include Adam and Adagrad. Different optimization tools and learning rates can affect a CNN's performance in compiling the image classification model. 

CNNs are susceptible to overfitting due to a large number of parameters in the system \citep{alzubaidi_review_2021}.
To mitigate this, techniques such as weight regularization ($\ell^1$, $\ell^2$) in convolutional layers, dropout in dense layers, and early stopping are widely employed. Specifically, random dropout can deactivate some neurons of a layer during training and is shown effective in image classifications \citep{garbin_dropout_2020}. 
Early stopping, which controls the number of training iterations (the epochs) to avoid overfitting when no improvement in parameter updates on a validation set, is simple yet effective
\citep{Anjali1713454}.

Our implementation of CNNs for brain tumor MRI image classification was conducted on TensorFlow Core v2.7.0 \citep{tensorflow2015_whitepaper} using Python 3.7 \citep{python10_5555_1593511}, containerized in Singularity \citep{si10_1371}. This was executed at the University of Arizona HPC, AMD Zen2 48 core processors with 18GB memory. We did not use NVIDIA CUDA GPU acceleration for fair comparison with other methods not utilizing parallel computing. In this paper, we implemented four distinct CNN architectures for brain tumor MRI image classification and summarized the results in Table \ref{table:table_1} \footnote{This table presents four distinct CNN architectures. BN: Batch Normalization, LR: Model Learning Rate. Batch normalization is systematically applied after each convolutional layer and prior to the pooling layer. A dropout layer is incorporated between adjacent layers within the dense network. The configuration of the hidden layer is denoted as $n$-tuples, where $n$ is the number of hidden layers, and each tuple element corresponds to the neuron count in that specific layer. The ReLU activation is consistently employed across both convolutional and dense networks in all configurations. Early stopping is implemented during the compilation of all models.}.  

\begin{table}[!htbp]
\caption{Various Configuration Settings of Four CNN Architectures}
\label{table:table_1}
\centering
\resizebox{\textwidth}{!}{
\begin{tabular}{c|c|c|c|c|c|c|c|c|c|c|c|c|c}  
\hline
\multirow{3}{*}{\textbf{CNN}~} & \multicolumn{9}{c|}{\textbf{Tumor MRI Image Feature Extraction}}                                                                                                                    & \multicolumn{2}{c|}{\textbf{Tumor Classification}} & \multicolumn{2}{c}{\multirow{2}{*}{\textbf{Model Compilation}}}  \\ 
\cline{2-12}
                               & \multicolumn{8}{c|}{\textbf{Convolutional Layers}}                                                                                            & \multirow{2}{*}{\textbf{Pooling~}} & \multicolumn{2}{c|}{\textbf{Dense Layer}}         & \multicolumn{2}{c}{}                                             \\ 
\cline{2-9}\cline{11-14}
                               & \textbf{Filter} & \textbf{Kernel~} & \textbf{Layers} & \textbf{Strides} & \textbf{Padding} & \textbf{L1} & \textbf{L2} & \textbf{BN} &                                    & \textbf{Hidden Layers} & \textbf{Dropout~}         & \textbf{Optimizer} & \textbf{LR}                                 \\ 
\hline
CNN1                           & 32              & 3                & 3               & $(1,1)$            & Valid            & 0           & 0           & No          & Max                                & $(255,)$                    & 0                         & Adam               & 0.001                                       \\
CNN2                           & 32              & 2                & 4               & $(2,2)$            & Same             & 0.01        & 0.001       & Yes         & Average                            & $(128,)$                    & 0.15                      & Adagrad            & 0.01                                        \\
CNN3                           & 64              & 4                & 5               & $(1,1)$            & Same             & 0.001       & 0.01        & Yes         & Average                            & $(512,128)$              & 0.5                       & Adagrad            & 0.001                                       \\
CNN4                           & 128             & 2                & 4               & $(2,2)$            & Same             & 0           & 0           & Yes         & Max                                & $(512,255,128)$          & 0.25                      & Adam               & 0.01                                        \\
\hline
\end{tabular}}
\end{table}

\subsection{Bag-of-Features (BoF) Model for Image Classification}\label{ssec:3.2}

Inspired by the Bag-of-Words model used extensively in natural language processing (NLP) and information retrieval, the Bag-of-Features (BoF) approach has been successfully applied to video retrieval \citep{Sivic2003} and image classification \citep{Csurka04}. This technique combines features, represented as $D$-dimensional vectors extracted from each normalized training image, into an unordered collection of features pool to form the BoF. This creates a comprehensive representation of the image dataset as a histogram of local descriptors. To reduce a typically large size of the BoF, stemming from numerous features and training images, a common approach is to cluster these local descriptor features using algorithms such as multi-pass $k$-means \citep{lutz_efficient_2018}. The resulting $M$-centroids form a $\realset^{M \times D}$ codebook (vocabulary) and offer a compact representation of the image's local descriptors. If hard vector encoding (VQ) is applied, each visual code has only one non-zero element; in soft-vector encoding (soft-VQ), a small subset of elements will be non-zero \citep{Wang5540018}. Consequently, each image is represented by multiple visual codes, encoded as an $\realset^{M \times S_i}$ matrix, where $S_i$ denotes the number of descriptors extracted from the $i$-th image. This image encoding matrix is then pooled and normalized to generate the final representation of the image features as a $M$-vector, which is equal to the codebook size, for downstream image classification. This process is carefully illustrated in Fig.\ref{fig:figure_1}.

\subsubsection{SIFT-based Image Feature Extraction}\label{ssec:3.2.1}
The goal of image feature extraction is to capture the most relevant and distinctive
characteristics of image patches. Local image descriptors play a pivotal role in tasks such as image classification and object detection in computer vision \citep{Fan10_5555_2898913}. For feature description, we employ the state-of-the-art Scale-Invariant Feature Transform (SIFT) technology \citep{lowe790410}, renowned for its robustness in extracting keypoints under image distortions, rotations, scaling, and noise. SIFT is particularly well-suited for extracting features from brain tumor MRI images, which vary in size, angle, and resolution. The SIFT algorithm utilizes cascade filters to compute the Difference of Gaussians (DoG) function progressively on rescaled images, detecting scale-invariant keypoints described by gradient orientation histograms, resulting in a 128-dimensional vector as the scale-invariant local descriptor. The algorithms are detailed in \citet{lowe_distinctive_2004}. For each image, SIFT calculates all the keypoints and the corresponding local descriptors, aggregating them to form the image feature representation. This produces a $\realset^{S_i \times D}$ extracted feature matrix, where $S_i$ is the count of keypoints detected by SIFT for the $i$-th image, and $D$ represents the dimension of the local keypoint descriptor, with $D=128$ in our case.

The original brain MRI images in the data set are in JPEG format and vary in size. We utilize MATLAB R2020b (The MathWorks, Natick, MA, USA) to uniformly resize each image to a resolution of $384 \times 384$ pixels, as discussed in \ref{ssec:imgprepro}, and then convert them to the PGM format for SIFT input. We run \texttt {sift.exe}, downloaded from the open-source VLFeat system at \url{https://www.vlfeat.org/install-shell.html}, to compute the SIFT features for the $i$-th image. The output is a text file, with each row comprising 128 numbers representing a SIFT keypoint descriptor (the initial four numbers, denoting X, Y coordinates, scale, and direction of each keypoint, are removed), and $S_i$ indicates the number of detected keypoints. 

\subsubsection{Codebook Generation}\label{ssec:3.2.2}
The Codebook is derived from the BoF feature pool and computed by the multi-pass $k$-means algorithm to establish a compact representation of image local descriptors. This is a two-step process outlined as follows. 

\paragraph*{(1) BoF Feature Pool} 
We combine the SIFT descriptors from all the training images to create the BoF feature pool as $\bigcup_i{S_i \times D}$, where $i$ is the index for each training image, $D$ is the SIFT descriptor dimension. Specifically, $D=128$. The BoF pool is typically a large matrix, and in our case, the matrix dimensions reach approximately $8\mathrm{e}6 \times 128$, making it infeasible for the downstream encoding process. 

\paragraph*{(2) Create the Codebook}
The BoF pool is aggregated to a pre-determined codebook size by multi-pass $k$-means \citep{lutz_efficient_2018}. $k$-means clustering is a well-established method for vector quantization and is widely employed in constructing codebooks or vocabularies \citep{Ibrahim9717942}. Its primary objective is to partition the set of $N$-descriptors into $M$ clusters, with each descriptor assigned to the cluster with the nearest centroid. These $M$-centroids, often referred to as \emph{visual words}, are assembled to create the $\realset^{M \times D}$ codebook as a compact representation of the image's local descriptors. The choice between 1-pass $k$-means and 2-pass $k$-means depends on the pre-defined codebook size and the available computation resources. In our case, we use 2-pass $k$-means when $M>128$ to ensure efficient computation.

\subsubsection{Feature Encoding}\label{ssec:3.2.3}
In the feature encoding phase, the SIFT local descriptor is converted to a ``\emph{visual code}", represented as an $M$-vector, by using the aggregated codebook described above. Common encoding methods include hard vector quantization (VQ) and soft vector quantization (soft-VQ). The key difference is that VQ has only one non-zero element for each visual code, while soft-VQ techniques like Locality-constrained Linear Coding (LLC) may involve a few non-zero elements \citep{Wang5540018}. In this paper, we implement both encoding schemes and compare their performance. The details are provided as follows.

\paragraph*{(1) Notations} 
Let $\bX = [\bx_1, \bx_2, \ldots, \bx_N]^{\small\mathrm{T}} \in \realset^{N \times D}$ be the set of local descriptors extracted from images using SIFT, $N$ is the number of descriptors, and $D$ is the dimension of each descriptor. Denote the codebook with $M$-centroids as $\bB = [\bb_1, \bb_2, \ldots, \bb_M]^{\small\mathrm{T}} \in \realset^{M \times D}$. During the encoding phase, each $D$-dimensional descriptor is converted into an $M$-dimensional visual code. The final image encoding is represented by a set of visual codes, $\bC = [\bc_1, \bc_2, \ldots, \bc_N]^{\small\mathrm{T}} \in \realset^{N \times M}$.

\paragraph*{(2) Encoding with VQ} 
The hard vector quantization (VQ) encoding solves the following constrained least squares optimization problem:
 \begin{mini}
   {\tiny \bC}{\sum_{i=1}^{N}\Vert\bx_i - \bB^{\small\mathrm{T}}\bc_i\Vert^2
  \label{opt1_vq}}{}{}
  \addConstraint{\Vert \bc_i \Vert_{0}}{= 1}, ~~ \Vert \bc_i \Vert_{1}= 1, ~~  {\bc_i}{\ge 0} ~~  \forall i.  
 \end{mini}
The $\ell^0$-norm constraint implies that the number of non-zero elements of $\bc_i$ is 1, while the $\ell^1$-norm and non-negative constraints guarantee that the non-zero element is equal to 1. This constraint-based approach identifies the nearest neighbor of a given descriptor among the $M$-centroids using the Euclidean distance and then assigns a value of $1$ in the visual code at the position corresponding to the nearest centroid in the codebook. The time complexity of this process is $O(M)$, which is linear in the size of the codebook.  

\paragraph*{(3) Encoding with Locality-constrained Linear Coding (LLC)} 
The soft-VQ encoding may use Lasso-regularized sparse coding with a relaxation of the $L_0$-norm constraint (SC-VQ), or add locality constraints using the LLC encoding. \citet{yu_nonlinear_2009} suggests that locality is more essential than sparsity, and \citet{Wang5540018} shows that the LLC encoding outperforms SC-VQ and possesses several favorable properties. In particular, the LLC encoding solves the following regularization  problem:
\begin{mini}
   {\tiny \bC}{\sum_{i=1}^{N}\Vert\bx_i - \bB^{\small\mathrm{T}}\bc_i\Vert^2 + \lambda\Vert\bd_i\odot\bc_i\Vert^2
  \label{opt2_llc}}{}{}
  \addConstraint{\mathbf{1}^{\small\mathrm{T}}\bc_i}{= 1}, ~~{ \forall i},
 \end{mini}
where $\odot$ represents the Hadamard product, and $\bd_i = \exp(\textbf{dist}(\bx_i,\bB)/\sigma) \in \realset^{M}$ stands for the locality adaptor, which measures the similarity of the input descriptor to each visual word in the codebook based on the Euclidean distance, and $\lambda$ and $\sigma$ are hyperparameters to be tuned \citep{Wang5540018}. The time complexity of this process is $O(M^2)$, which could be computationally inefficient when the number of entries in the codebook is large. 

\paragraph*{(4) Approximate Encoding with Fast LLC}
The solution to \ref{opt2_llc} has few significant values, which in fact performs feature selection of local bases for each descriptor. In practice, only those bases proximate to the given descriptor have non-zero coefficients. Therefore, we can develop a fast approximation of LLC by simply computing the $K$-nearest neighbors within the codebook for a given $i$-th descriptor. This results in a reduced local base system, denoted as $\bB_i \in \realset^{K \times D}$, where typically $K<D<M$. Subsequently, we can solve a much smaller constrained linear system for the visual code:
\begin{mini}
   {\tiny \tilde{\bC}}{\sum_{i=1}^{N}\Vert\bx_i - \bB_i^{\small\mathrm{T}}\tilde{\bc}_i\Vert^2
  \label{opt3_fllc}}{}{}
  \addConstraint{\mathbf{1}^{\small\mathrm{T}}\tilde{\bc}_i}{= 1},~~{ \forall i},
 \end{mini}
where $\tilde{\bC} = [\tilde{\bc}_1, \tilde{\bc}_2, \ldots, \tilde{\bc}_N]^{\small\mathrm{T}} \in \realset^{N \times K}$. The complete code can be generated by placing the $K$ elements of $\tilde{\bc}_i$ in the correct positions within $\bc_i$ while setting the remaining $M-K$ elements to 0. The fast LLC encoding reduces the time complexity to $O(M+K^2)$, given that $K \ll M$, achieves a linear complexity of $O(M)$, which is the same as VQ. However, in practice, VQ is still much faster since LLC encoding typically requires a larger codebook, such as $M>1024$, to obtain optimal results \citep{Wang5540018}.

\subsubsection{Pooling and Normalization}\label{ssec:3.2.4}
The image is now represented as a collection of encoding vectors $\bC \in \realset^{N \times M}$, where each descriptor is mapped to a given codebook. These vectors need to be aggregated into a single vector $\bq \in \realset^{M}$ as an input for downstream image classification. This is done by pooling the image encoding features as $\tilde{\bq} \in \realset^{M}$, which are then normalized to obtain the final image feature representation as the $M$-vector $\bq$. The details are described as follows.
 
\paragraph*{(1) Pooling} We can do sum pooling or max pooling:
\hfill \break 
\hspace*{1.5em}
\textbf{A.} Sum pooling: $\tilde{q}_j = \sum_{i=1}^{N}\bC_{ij}$, $j = 1,2,\ldots, M$
\hfill \break \hspace*{1.5em}
\textbf{B.} Max pooling: $\tilde{q}_j = \max_{i=1}^{N}\bC_{ij}$, $j = 1,2,\ldots, M$
\hfill \break 
In the case of VQ encoding, the sum pooling computes the histogram of the number of occurrences of each visual code in images, while the max pooling generates a binary histogram indicating the presence or absence of each visual code in images. 

\paragraph*{(2) Vector Normalization}
Our work shows that various normalization methods for the pooling vector $\tilde\bq$ have a significant impact on downstream image classification. We employ the following three normalization methods:
\hfill \break \hspace*{1.5em}
\textbf{A.} Sum normalization: $\bq = \tilde{\bq}/\sum_j\tilde{q}_j$, $j = 1,2,\ldots, M$
\hfill \break \hspace*{1.5em}
\textbf{B.} L2 normalization: $\bq = \tilde{\bq}/\Vert\tilde{\bq}\Vert_2$
\hfill \break \hspace*{1.5em}
\textbf{C.} Log term frequency weight (LTF) normalization: 
\hfill \break \hspace*{1.5em}
We develop this normalization method inspired by the log term frequency weighting within the vector space model of information retrieval theory \citep{manning_2008}. In this approach, we view each element of the pooling vector $\tilde{q}_j$ as the term frequency within a document containing $M$ terms. The LTF normalization process is different for pooling from VQ and LLC, as encoding from LLC may include negative values in the pooling vector. For each $j = 1,2,\ldots, M$, we obtain the normalized vector $\bq$ from the pooling vector $\tilde{\bq}$ as follows:
\hfill \break \hspace*{3em}
\textbf{(a).} Pooling from VQ encoding
\[
    q_j= 
\begin{cases}
    1 + \log_{10}{\tilde{q}_j} & \text{if } \tilde{q}_j > 0\\
    0               & \text{otherwise}
\end{cases}
\]
\hfill \break \hspace*{3em}
\textbf{(b).} Pooling from LLC encoding involves the following steps: (i) set all the negative term frequencies of $\tilde{q}_j$ to 0; (ii) calculate the minimum positive term frequency, defined as $\tilde{q}_{min} = \min_j{\tilde{q}_j}$, where $\tilde{q}_j > 0$; (iii) perform scaling $\tilde{\bq}^* = \tilde{\bq}/\tilde{q}_{min}$; (iv) use $\tilde{\bq}^*$ as input and follow the LTF normalization procedure as VQ in (a).

Different combinations of pooling and normalization methods can be used to generate the final image vector. These choices can lead to significant differences in performance for downstream classification within the wSVM framework, as elaborated in \ref{sec:res4}. 

\subsection{Class Probability Estimation via Weighted SVMs}\label{ssec:3.3}
Once we obtain the vector representation of images as $\bq \in \realset^{M}$, the final step is to perform multiclass classification and probability estimation. Towards this, we employ the weighted Support Vector Machines (wSVMs), which are designed for estimating posterior class probabilities in $K$-class problems through ensemble learning \citep{WuZhaLiu2010, wang_multiclass_2019,zeng_wsvms_2022}. The method is entirely non-parametric, and hence both flexible and robust. Given any input $\bx$, we estimate its class
probabilities $\hat{p}_j(\bx)$ for $j=1,\ldots, K$, and then construct the classifier by the argmax rule, $\hat{\phi}(\bx)= \arg\max_{1 \le j \le K}\hat{p}_j(\bx)$, or using the max voting algorithm based on the full pairwise conditional probability table \citep{zeng_wsvms_2022}. 

Various learning schemes have been proposed for multiclass probability estimation within the wSVMs framework. \citet{WuZhaLiu2010} developed an algorithm that simultaneously learns multiple probability functions, which has nice theoretical properties and empirical performance. However, it involves exponential time complexity. \citet{wang_multiclass_2019} proposed the pairwise coupling technique based on divide-and-conquer, reducing the complexity from exponential to polynomial in $K$. Recently, \citet{zeng_wsvms_2022} introduced the baseline learning scheme to reduce the computational cost, achieving a linear complexity in $K$ and making the wSVMs scalable for massive data with a large $K$. Furthermore, baseline learning can enhance the pairwise coupling, improving both computation time and estimation accuracy. Finally, the One-vs-All (OVA) learning scheme developed by \citet{zeng_wsvms_2022} exhibits the best empirical performance, although it is not the most computationally efficient. In the following section, we start with a brief review of the wSVM probability estimation framework.

\paragraph*{(1) Notations}
Denote the entire labeled training set as $\left\{\left(\bx_i,y_i\right), i=1,\ldots,n\right\}$, where $\bx_i=(x_{i1}, \ldots, x_{ip})^{\small\mathrm{T}}\in\mathbb{R}^p$, $n$ is the training sample size, and $p$ is the data dimension. For each class $j=1, \cdots, K$, define its training set as
$\mathcal{S}_j = \{(\bx_i, y_i) \mid y_i = j; ~~ i=1, \ldots, n\}$. The estimated class probability functions from the training data are denoted by $\boldsymbol{\hat{p}}= \{\hat{p}_1,\ldots,\hat{p}_K\}$, where the magnitude of $\hat{p}_j$ signifies the confidence level of a data point belonging to class $j$. The classification boundary is determined by either the argmax rule, $\hat{\phi}(\bx)= \arg\max_{1 \le j \le K}\hat{p}_j(\bx)$, or the max voting algorithm \citep{zeng_wsvms_2022} if the pairwise conditional probability table is available.

\paragraph*{(2) Binary wSVMs}
In binary classification problems, the class label $y\in\{+1, -1\}$. For each input $\bx$, we define its posterior class probabilities as $p_{+1}(\bx)=P(Y=+1|\bX=\bx)$ and $p_{-1}(\bx)=P(Y=-1|\bX=\bx)$, respectively. The binary wSVM solves the following optimization problem:
        \begin{equation}
          \min_{f\in\mathcal{H}_K}\frac{1}{n}\Big[(1-\pi)\sum_{y_i=1}L({y_if(\bx_i)})+\pi\sum_{y_i=-1}L({y_if(\bx_i)})\Big]+\lambda J(f),
            \label{wsvm1}
        \end{equation}
where $L(yf(\bx))=(1-yf(\bx))_+=\max\{0, 1-yf(\bx)\}$ is the hinge loss function, $\mathcal{F}$ is some functional space, the regularization term $J(f)$ controls model complexity, and the parameter $\lambda>0$ balances the bias–variance trade-off \citep{Hastie2009}. For kernel SVMs, $f$ employs a bivariate Mercer kernel representation $\mathbf{K}(\cdot, \cdot)$ of the form $f(\bx)=d +\sum_{i=1}^n c_i\mathbf{K}(\bx_i, \bx)$ according to the representer theorem \citep{KW1971}, and the space $\mathcal{F}=\mathcal{H}_\mathbf{K}$ is the reproducing kernel Hilbert space (RKHS) induced by $\mathbf{K}(\cdot,\cdot)$ \citep{Wahba90}, and $J(f)=\norm1[\big]{\bm{f}}_{\mathcal{H}_\mathbf{K}}^2=\sum_{i=1}^n\sum_{j=1}^{n}c_ic_j\mathbf{K}(\bx_i, \bx_j)$. It is known that, for each weight $\pi$, the minimizer of the expected weighted hinge loss $\mathop{\mathbb{E}}_{(\bX,Y)\sim P(\bX, Y)}\left\{W(Y)L[Yf(\bX)]\right\}$ has the same sign as sign[$p_{+1}(\bX)-\pi$] \citep{lin_support_2002}. To estimate posterior class probabilities, we first train multiple classifiers $\hat{f}_{\pi_1},\cdots, \hat{f}_{\pi_{M}}$ by solving \eqref{wsvm1} with a series of $\pi$ values satisfying $0<\pi_1<\cdots<\pi_{M}<1$. Theoretically speaking, for each input $\bx$, the sequence $\hat{f}_\pi(\bx)$ is non-increasing in $\pi$, resulting a unique value $m^*$ such that $\pi_{m^*}<p_{+1}(\bx)<\pi_{m^*+1}$. A consistent binary class probability estimator can then be constructed as
$\hat{p}_{+1}(\bx)=\frac{1}{2}(\pi_{m^*}+\pi_{m^*+1})$.

\paragraph*{(3) Multiclass wSVMs}
For $K$-class problems, where the class label $y\in\{1, 2, \ldots, K\}$ with $K\ge 3$, the goal is to estimate class probabilities $p_j(\bx)=P(Y=j|\bX=\bx)$ for $j=\{1,\ldots, K\}$. When considering any pair of two classes, denoted as $j$ and $j'$ with $1 \le j \neq j' \le K $, we define the pairwise posterior conditional probability of a data point belonging to class $j$, given that it must belong to either class $j'$ or class $j$, as
\begin{equation}\label{conprob}
q_{j|(j, j')}(\bx) = \frac{P(Y=j|\bX=\bx)} {\sum_{i \in \{j,j'\}} P(Y=i|\bX=\bx)} = \frac{p_j(\bx)}{p_j(\bx) + p_{j'}(\bx)}.
\end{equation}

\citet{wang_multiclass_2019} proposed the pairwise coupling method, which decomposes the $K$-class problem into $\binom{K}{2}$ binary problems and constructs the class probabilities using all pairwise conditional probability as follows
\begin{equation}
            p_{j}(\bx)=\frac{ q_{j|(j, k)}(\bx)/ q_{k|(j,k)}(\bx)}{\sum_{s=1}^{K} q_{s|(s, k)}(\bx)/ q_{k|(s,k)}(\bx)}, j=1,\cdots, K; ~~ k \neq j. \label{allprob1}
\end{equation}
Each pairwise conditional probability is estimated using binary wSVMs. \citet{zeng_wsvms_2022} proposed an algorithm with linear time complexity by selecting a fixed common baseline class $k^*$ to train $K-1$ binary wSVMs problems. Alternatively, one can use the One-vs-All (OVA) approach, which involves training $K$ binary wSVMs problems for classifying $j$ vs ``not j" for $j=1,\ldots, K$. 

In this work, different codebook sizes and encoding methods may introduce varying levels of complexity in multiclass probability estimation using the wSVMs. We employ the following implementations for the brain tumor MRI image analysis:  (i) the pairwise coupling method, denoted as SIBOW-PSVM; (ii) the OVA learning scheme, denoted as SIBOW-ASVM; (iii) the baseline learning scheme, denoted as SIBOW-BSVM, which selects the baseline class either based on the abundance of class data or by considering the aggregated class distance, along with pairwise probability reconstruction as proposed in \citet{zeng_wsvms_2022}, denoted as SIBOW-BPSVM. It is observed that the two baseline methods provide similar results.

\subsection{Model Calibration}\label{ssec:3.4}
To measure the accuracy of multiclass probability estimation across different algorithms, particularly in the absence of true underlying probabilities, we employ model calibration to evaluate the correctness of probability estimation \citep{guo2017}. Specifically, to estimate the expected accuracy of class probabilities, we first group the predicted class probabilities $\hat{p}_{\hat{y}_i}(\tilde \bx_i)$ into $M$ interval bins, each with a size of $\frac{1}{M}$. Define $P_m = \{i \mid \hat{p}_{\hat{y}_i}(\tilde \bx_i) \in ({\frac{m-1}{M},\frac{m}{M}}]\}$ for $m = 1,\ldots, M$ and $i = 1, \ldots,\tilde{n}$. We then calculate the accuracy of $P_m$ as: 
$\text{acc}(P_m) = \sum_{i \in P_m}I(\hat{y}_i = \tilde {y}_i)/\Card{P_m},$
where $\hat{y}_i$ and $\tilde{y}_i$ are the predicted and true class labels for the $i$-th test sample, and $I(\cdot)$ is the indicator function. The estimated confidence within bin $P_m$ is defined as: 
$\text{conf}(P_m) = \sum_{i \in P_m}\hat{p}_{\hat{y}_i}(\tilde\bx_i)/\Card{P_m}.$ 
To visualize model calibration, we employ the {\it reliability diagram} tool, which graphically represents the expected sample accuracy as a function of confidence. In a perfectly calibrated model, we would have $\text{acc}(P_m) = \text{conf}(P_m)$ for $\forall m \in 1,\ldots M$, resulting in a diagonal line at ${45}\degree$ on the reliability diagram. Any derivations from the ideal scenario indicate miscalibration \citep{Naeini2015ObtainingWC} at some level. The reliability curve for each method is smoothed using Generalized Additive Models (GAM) \citep{Hastie2009} and compared against the ${45}\degree$ line. We quantify the miscalibration in terms of the Expected Calibration Error (ECE), which measures the absolute difference between predictive confidence and accuracy,
$\text{ECE} = \sum_{m=1}^{M} \Card{P_m}\times \Card{\text{acc}(P_m)-\text{conf}(P_m)}/\tilde{n},$
where $\tilde{n}$ is the number of test samples. The smaller the ECE value, the better model calibration.

\section{Results and Discussion}
\label{sec:res4}
This section reports the classification results for the brain tumor MRI image data set, as described in \ref{ssec:imgprepro}. We compare the proposed SIBOW-SVM with the standard Convolutional Neural Networks (CNNs), in terms of both classification accuracy and probability estimation. The performance is evaluated based on the test set comprising 196 glioma, 185 meningioma, 173 pituitary, and 104 normal images with mixed planes. In addition to the misclassification error, we report the macro (average) precision, recall, and F-1 scores by treating all the classes equally. 

\subsection{CNNs}\label{ssec:4.1}
In Table \ref{table:table2}.\, \footnote{The table shows performance measurements for four CNN architectures described in \ref{ssec:3.1}. The running time is measured in minutes. The values for TE (misclassification rate), Precision, Recall, F1 (macro average), and ECE (expected calibration error) are multiplied by 100 for both mean and standard derivation (shown in parentheses). Training epochs were stopped early if the evaluation-loss increased.} We summarize the results of four CNN network architectures.  These CNNs exhibit different levels of complexity, which impact their image classification performance and computational efficiency. \blue{The computational complexity and the number of parameters of a CNN algorithm increases with the number of layers, filters, kernels, and image sizes. In our examples, CNN1 has the lowest complexity but contains more than 100 million parameters.} We randomly split the images into training and validation sets of equal sizes, repeated 10 times, and reported the average misclassification rates, macro precision/recall/F-1 scores, and ECE evaluated on the test set, along with standard errors. CNN1 generally gives the best classification and probability estimation performance, CNN4 has the overall worst performance, and CNN3 has the worst time complexity among the models. These results align with the Kaggle code pool with standard CNN implementations. 

\begin{table}[!htbp]
\caption{Performance measures on various CNN architectures.}
\label{table:table2}
\centering
\setlength{\tabcolsep}{3pt}
\scalebox{0.95}{
\begin{tabular}{c|cccc} 
\hline
\textbf{Metrics} & \textbf{CNN1} & \textbf{CNN2} & \textbf{CNN3} & \textbf{CNN4}  \\ 
\hline
Time             & 25.4 (1.8)    & 19.3 (0.7)    & 195.5 (4.2)   & 42.7 (2.8)     \\ 
\hline
TE               & 22.7 (0.4)    & 25.4 (0.6)    & 24.1 (0.7)    & 29.2 (0.8)     \\ 
\hline
Precision        & 76.9 (0.4)    & 74.5 (0.9)    & 76.4 (0.7)    & 70.8 (0.8)     \\
Recall           & 78.3 (0.4)    & 76.4 (0.7)    & 75.9 (0.5)    & 75.4 (1.1)     \\
F1               & 76.7 (0.4)    & 73.7 (0.7)    & 75.1 (0.9)    & 70.8 (0.9)     \\ 
\hline
ECE              & 20.2 (0.3)    & 20.5 (0.3)    & 20.4 (0.3)    & 21.0 (0.7)     \\ 
\hline
Epoches          & 7.7 (0.4)     & 41.6 (1.3)    & 19.8 (0.4)    & 17.4 (1.1)     \\
\hline
\end{tabular}}
\end{table}

\subsection{The Proposed SIBOW-SVM Model}\label{ssec:4.2}
This section presents the results of the proposed SIBOW-SVM model for image classification and compares them with those given by the standard CNN architecture. \blue{The total number of parameters involved in training the weighted-SVM classifier depends on the choice of kernels. If a linear kernel is used, the SIBOW-SVM is a linear classifier with $M$ parameters, where $M$ is the number of centroids during $k$-means in the codebook generation phase. If a radial basis function (RBF) kernel is used, the SIBOW-SVM is fully nonparametric and becomes a nonlinear classifier, and its solution is obtained by solving its dual problem, which involves $n_i$ parameters, where $n_i$ is the number of input images, also known as the training sample size. We solve the SIBOW-SVM optimization problem by quadratic programming (QP), which has computational complexity as $O(Kn_i^3)$, where $K$ is the number of classes \citep{zeng_wsvms_2022}. The complexity of a 2-dimensional CNN is $\simeq O(n_p^2)$,  where $n_p$ is the total number of pixels of the input images including all channels \citep{oyedare2023keep}. For reasonably-sized RGB images, $Kn_i^3 \ll n_p^2$. As such, SIBOW-SVM presents significantly lower complexity than CNNs.} 

We extracted SIFT features uniformly as a BoF pool and implemented two feature encoding methods: VQ and LLC, as detailed in Section \ref{ssec:3.2}. These methods require different codebook generation approaches. All encoding features were then pooled and normalized, for example, using ``sum pooling" followed by ``L2 normalization", referred to as ``sum-L2" pooling.  In total, we employed six pooling methods, as outlined in \ref{ssec:3.2.4}, denoted as (1)  ``sum-sum'', (2) ``sum-L2'', (3) ``sum-LTF'', (4)  ``max-sum'', (5) ``max-L2'', and (6) ``max-LTF'', as the feature inputs for the wSVMs.


Pooling methods play a critical role in the performance of wSVMs. For VQ encoding, we observe a significant performance difference between two pooling methods: ``sum-LTF", which uses TF-IDF weighting inspired by information retrieval; and ``sum-sum", which employs frequency pooling. Some pooling combinations yielded similar results, and we report the best results.

For kernel wSVMs, hyperparameters were tuned using EGKL, following the procedure in \citet{zeng_wsvms_2022}. We randomly split the images into the training and tuning sets of equal sizes, repeated 10 times, and reported the average misclassification rates, macro precision/recall/F-1 scores, and ECE evaluated on the test image set, along with standard errors.

\subsubsection{VQ Encoding Scheme}\label{ssec:4.2.1}
The VQ encoding generally requires a small codebook size due to its nature of hard vector quantization \citep{AdamCoatesN11,wu_vector_2019}, making it more computationally efficient for complex problems as we can perform 1-pass $k$-means in most cases. In our experiments, we implemented codebooks with sizes of $M \in \{32, 64, 128\}$.  

\paragraph*{Pooling with ``sum-LTF"} Table \ref{table:table_3}\footnote{The table presents performance measurements for VQ encoding with sum-LTF pooling using three codebook sizes. The running time is measured in minutes. TE1 (misclassification rate based on \emph{max probability}), TE2 (misclassification rate based on \emph{max voting}), macro precision, macro recall, macro F1, and ECE (expected calibration error) are multiplied by 100 for both mean and standard derivation (shown in parentheses). The optimal baseline class $k^*$ for the baseline methods is calculated using the statistical mode \citep{zeng_wsvms_2022}. Similar explanations apply to the results for other encoding methods.} reports the VQ encoding with the newly developed ``sum-LTF'' pooling method. We denote schemes with different codebook sizes as \textbf{VQ32}, \textbf{VQ64}, and \textbf{VQ128}. Notably, the VQ64 encoding with wSVMs enhanced by pairwise coupling and baseline learning (BP1-SVM and BP2-SVM) yielded the best performance, showing a 12\% increase in classification accuracy, a 14\% increase in macro-precision, a 12\% increase in macro-recall, and a 15\% increase in macro-F1 score.  For multiclass probability estimation, it resulted in a 57\% reduction in ECE (model calibration error) and a 63\% reduction in computation time compared to the standard convolutional neural network's best performer CNN1. The OVA-SVM (A-SVM) consistently delivered the most accurate estimation. However, it required longer computation times. The codebook size was critical, as the performance decreased when $M$ changed, as shown in the VQ32 and VQ128 schemes.

\paragraph*{Pooling with ``sum-sum"} Table \ref{table:table_4} reports the results for VQ encoding with ``sum-sum'' as frequency pooling. We denote schemes with different codebook sizes as \textbf{PF32}, \textbf{PF64}, and \textbf{PF128}. It was observed that, with the same encoding vector, changing the pooling method from ``sum-LTF" to ``sum-sum" led to a significant decrease in performance. PF128 was the best performer, showing a 15\% decrease in classification accuracy, a 17\% decrease in macro F1, and a 94\% increase in ECE. It provided the optimal results only with the newly developed ``sum-LTF" pooling.

\begin{table}[!htbp]
\caption{Performance measure of VQ encoding with sum-LTF pooling}
\label{table:table_3}
\centering
\setlength{\tabcolsep}{3pt}
\scalebox{0.75}{
\begin{tabular}{c|cc|cc|c|c} 
\hline
\textbf{VQ32}  & \textbf{B1-SVM} & \textbf{BP1-SVM} & \textbf{B2-SVM} & \textbf{BP2-SVM} & \textbf{A-SVM} & \textbf{P-SVM}  \\ 
\hline
Time           & 13.1 (0.2)      & 13.1 (0.2)       & 11.9 (0.8)      & 11.9 (0.8)       & 143.1 (3.2)    & 22.4 (0.2)      \\ 
\hline
TE1            & 19.1 (0.4)      & 18.9 (0.3)       & 19.1 (0.4)      & 19.0 (0.4)       & 18.3 (0.3)     & 19.0 (0.4)      \\
TE2            & NA (NA)         & 18.7 (0.3)       & NA (NA)         & 18.8 (0.4)       & NA (NA)        & 18.8 (0.3)      \\ 
\hline
Precision      & 81.8 (0.2)      & 81.9 (0.2)       & 81.5 (0.2)      & 81.8 (0.3)       & 82.2 (0.4)     & 81.7 (0.4)      \\
Recall         & 81.4 (0.3)      & 81.6 (0.2)       & 81.5 (0.3)      & 81.6 (0.3)       & 82.8 (0.4)     & 81.6 (0.4)      \\
F1             & 81.2 (0.3)      & 81.4 (0.2)       & 81.2 (0.3)      & 81.3 (0.3)       & 82.2 (0.4)     & 81.3 (0.4)      \\ 
\hline
ECE            & 12.4 (0.4)      & 12.2 (0.3)       & 12.5 (0.4)      & 12.3 (0.3)       & 11.8 (0.2)     & 12.3 (0.5)      \\ 
\hline
$k^*$     & 2               & 2                & 4               & 4                & NA (NA)        & NA (NA)         \\ 
\hline
\textbf{VQ64}  & \textbf{B1-SVM} & \textbf{BP1-SVM} & \textbf{B2-SVM} & \textbf{BP2-SVM} & \textbf{A-SVM} & \textbf{P-SVM}  \\ 
\hline
Time           & 11.0 (0.1)      & 11.0 (0.1)       & 9.5 (0.7)       & 9.5 (0.7)        & 107.6 (2.0)    & 19.4 (0.2)      \\ 
\hline
TE1            & 14.3 (0.2)      & 14.3 (0.2)       & 14.4 (0.2)      & 14.3 (0.2)       & 13.7 (0.2)     & 14.3 (0.2)      \\
TE2            & NA (NA)         & 13.8 (0.2)       & NA (NA)         & 13.9 (0.2)       & NA (NA)        & 13.9 (0.2)      \\ 
\hline
Precision      & 87.3 (0.2)      & 87.6 (0.1)       & 87.1 (0.2)      & 87.3 (0.2)       & 88.3 (0.2)     & 87.3 (0.1)      \\
Recall         & 86.9 (0.2)      & 87.2 (0.2)       & 86.7 (0.1)      & 86.9 (0.1)       & 87.7 (0.2)     & 86.8 (0.2)      \\
F1             & 87.2 (0.2)      & 87.5 (0.2)       & 87.1 (0.2)      & 87.2 (0.2)       & 88.0 (0.2)     & 87.1 (0.2)      \\ 
\hline
ECE            & 9.3 (0.3)       & 8.9 (0.3)        & 9.4 (0.3)       & 9.0 (0.2)        & 8.7 (0.3)      & 9.3 (0.2)       \\ 
\hline
$k^*$     & 2               & 2                & 2               & 2                & NA (NA)        & NA (NA)         \\ 
\hline
\textbf{VQ128} & \textbf{B1-SVM} & \textbf{BP1-SVM} & \textbf{B2-SVM} & \textbf{BP2-SVM} & \textbf{A-SVM} & \textbf{P-SVM}  \\ 
\hline
Time           & 10.7 (0.1)      & 10.7 (0.1)       & 8.5 (0.8)       & 8.5 (0.8)        & 97.7 (1.2)     & 18.5 (0.0)      \\ 
\hline
TE1            & 29.1 (0.4)      & 29.1 (0.4)       & 29.2 (0.4)      & 29.2 (0.4)       & 28.2 (0.3)     & 29.1 (0.5)      \\
TE2            & NA (NA)         & 29.1 (0.4)       & NA (NA)         & 29.1 (0.5)       & NA (NA)        & 29.1 (0.4)      \\ 
\hline
Precision      & 67.2 (0.4)      & 67.5 (0.3)       & 67.1 (0.2)      & 67.4 (0.2)       & 68.3 (0.5)     & 67.3 (0.4)      \\
Recall         & 83.3 (0.3)      & 83.5 (0.2)       & 83.2 (0.2)      & 83.4 (0.2)       & 84.2 (0.6)     & 83.4 (0.4)      \\
F1             & 68.8 (0.6)      & 69.1 (0.6)       & 68.8 (0.6)      & 69.0 (0.6)       & 70.1 (0.5)     & 69.0 (0.5)      \\ 
\hline
ECE            & 14.6 (0.4)      & 14.4 (0.3)       & 14.5 (0.4)      & 14.3 (0.2)       & 14.0 (0.5)     & 14.4 (0.5)      \\ 
\hline
$k^*$      & 2               & 2                & 3               & 3                & NA (NA)        & NA (NA)         \\
\hline 
\end{tabular}}
\end{table}

\begin{table}[!htbp]
\caption{Performance measure of VQ encoding with sum-sum pooling}
\label{table:table_4}
\centering
\setlength{\tabcolsep}{3pt}
\scalebox{0.75}{
\begin{tabular}{c|cc|cc|c|c} 
\hline
\textbf{PF32}  & \textbf{B1-SVM} & \textbf{BP1-SVM} & \textbf{B2-SVM} & \textbf{BP2-SVM} & \textbf{A-SVM} & \textbf{P-SVM}  \\ 
\hline
Time           & 17.6 (0.3)      & 17.6 (0.3)       & 16.5 (0.9)      & 16.5 (0.9)       & 247.0 (4.6)    & 29.9 (0.8)      \\ 
\hline
TE1            & 30.2 (0.3)      & 30.2 (0.3)       & 30.3 (0.3)      & 30.3 (0.3)       & 29.5 (0.2)     & 30.0 (0.3)      \\
TE2            & NA (NA)         & 29.8 (0.3)       & NA (NA)         & 29.9 (0.3)       & NA (NA)        & 29.8 (0.3)      \\ 
\hline
Precision      & 69.0 (0.3)      & 69.3 (0.4)       & 68.8 (0.3)      & 69.1 (0.3)       & 69.8 (0.2)     & 69.1 (0.3)      \\
Recall         & 70.6 (0.2)      & 70.9 (0.2)       & 70.5 (0.3)      & 70.7 (0.4)       & 71.9 (0.3)     & 70.8 (0.3)      \\
F1             & 69.3 (0.4)      & 69.6 (0.4)       & 69.2 (0.3)      & 69.5 (0.3)       & 70.5 (0.2)     & 69.5 (0.3)      \\ 
\hline
ECE            & 16.7 (0.2)      & 16.4 (0.2)       & 16.6 (0.2)      & 16.3 (0.2)       & 16.1 (0.3)     & 16.3 (0.3)      \\ 
\hline
$k^*$      & 2               & 2                & 4               & 4                & NA (NA)        & NA (NA)         \\ 
\hline
\textbf{PF64}  & \textbf{B1-SVM} & \textbf{BP1-SVM} & \textbf{B2-SVM} & \textbf{BP2-SVM} & \textbf{A-SVM} & \textbf{P-SVM}  \\ 
\hline
Time           & 17.2 (0.5)      & 17.2 (0.5)       & 17.3 (0.5)      & 17.3 (0.5)       & 235.8 (6.0)    & 29.7 (0.7)      \\ 
\hline
TE1            & 28.2 (0.1)      & 28.1 (0.2)       & 28.1 (0.1)      & 28.0 (0.1)       & 27.4 (0.2)     & 28.1 (0.2)      \\
TE2            & NA (NA)         & 27.8 (0.1)       & NA (NA)         & 27.8 (0.1)       & NA (NA)        & 28.0 (0.2)      \\ 
\hline
Precision      & 69.6 (0.5)      & 69.9 (0.4)       & 69.9 (0.4)      & 71.1 (0.4)       & 71.6 (0.3)     & 71.0 (0.2)      \\
Recall         & 72.7 (0.3)      & 72.8 (0.4)       & 72.6 (0.3)      & 72.8 (0.2)       & 73.4 (0.3)     & 72.7 (0.2)      \\
F1             & 71.3 (0.5)      & 71.4 (0.4)       & 71.5 (0.4)      & 71.6 (0.4)       & 72.4 (0.3)     & 71.5 (0.2)      \\ 
\hline
ECE            & 16.7 (0.3)      & 16.4 (0.3)       & 16.9 (0.3)      & 16.6 (0.3)       & 16.0 (0.2)     & 16.6 (0.2)      \\ 
\hline
$k^*$          & 1               & 1                & 2               & 2                & NA (NA)        & NA (NA)         \\ 
\hline
\textbf{PF128} & \textbf{B1-SVM} & \textbf{BP1-SVM} & \textbf{B2-SVM} & \textbf{BP2-SVM} & \textbf{A-SVM} & \textbf{P-SVM}  \\ 
\hline
Time           & 16.3 (0.3)      & 16.3 (0.3)       & 14.6 (1.0)      & 14.6 (1.0)       & 228.6 (5.7)    & 28.0 (0.4)      \\ 
\hline
TE1            & 26.7 (0.2)      & 26.7 (0.2)       & 26.7 (0.2)      & 26.6 (0.2)       & 25.9 (0.2)     & 26.6 (0.1)      \\
TE2            & NA (NA)         & 26.1 (0.2)       & NA (NA)         & 26.1 (0.2)       & NA (NA)        & 26.4 (0.2)      \\ 
\hline
Precision      & 72.3 (0.3)      & 72.6 (0.4)       & 72.2 (0.4)      & 72.4 (0.4)       & 72.6 (0.2)     & 72.4 (0.2)      \\
Recall         & 74.5 (0.4)      & 74.8 (0.4)       & 74.3 (0.4)      & 74.6 (0.3)       & 75.4 (0.2)     & 74.7 (0.2)      \\
F1             & 73.0 (0.4)      & 73.1 (0.4)       & 72.8 (0.4)      & 72.9 (0.4)       & 73.6 (0.2)     & 73.0 (0.2)      \\ 
\hline
ECE            & 17.4 (0.2)      & 17.1 (0.2)       & 17.6 (0.4)      & 17.2 (0.4)       & 16.8 (0.3)     & 17.3 (0.4)      \\ 
\hline
$k^*$      & 2               & 2                & 1               & 1                & NA (NA)        & NA (NA)         \\
\hline
\end{tabular}}
\end{table}

\subsubsection{LLC Encoding Scheme}\label{ssec:4.2.2}
Compared to VQ, LLC encoding requires larger codebook sizes for optimal performance \citep{Wang5540018}. In our experiments, we implemented codebooks with sizes of $M \in \{2048, 4096,8192\}$. The performance for $M > 2^{13}$ drastically deteriorated in both computational efficiency and performance. Since $D \ll M$, we implemented the fast approximate LLC encoding described in Section \ref{ssec:3.2.3}. We chose $K = 5$, following \citet{Wang5540018}. We evaluated 6 pooling methods. 

\begin{table}[!htbp]
\caption{Performance measure of LLC encoding with max-L2 pooling}
\label{table:table_5}
\centering
\setlength{\tabcolsep}{3pt}
\scalebox{0.75}{
\begin{tabular}{c|cc|cc|c|c} 
\hline
\textbf{LLC2048} & \textbf{B1-SVM} & \textbf{BP1-SVM} & \textbf{B2-SVM} & \textbf{BP2-SVM} & \textbf{A-SVM} & \textbf{P-SVM}  \\ 
\hline
Time             & 34.8 (2.4)      & 34.8 (2.4)       & 34.3 (2.3)      & 34.3 (2.3)       & 252.9 (9.6)    & 60.8 (4.0)      \\ 
\hline
TE1              & 33.5 (0.6)      & 33.5 (0.6)       & 33.7 (0.6)      & 33.7 (0.6)       & 32.7 (0.6)     & 33.6 (0.6)      \\
TE2              & NA (NA)         & 33.2 (0.6)       & NA (NA)         & 33.3 (0.7)       & NA (NA)        & 33.3 (0.6)      \\ 
\hline
Precision        & 66.8 (0.7)      & 67.1 (0.7)       & 66.4 (0.6)      & 66.8 (0.6)       & 67.3 (0.5)     & 66.9 (0.5)      \\
Recall           & 68.8 (0.4)      & 69.0 (0.4)       & 68.6 (0.5)      & 68.9 (0.5)       & 69.1 (0.6)     & 69.0 (0.6)      \\
F1               & 67.3 (0.6)      & 67.6 (0.6)       & 67.1 (0.5)      & 67.4 (0.5)       & 68.1 (0.5)     & 67.5 (0.6)      \\ 
\hline
ECE              & 17.6 (0.4)      & 17.4 (0.3)       & 17.6 (0.2)      & 17.4 (0.3)       & 17.2 (0.4)     & 17.5 (0.3)      \\ 
\hline
$k^*$        & 2               & 2                & 1               & 1                & NA (NA)        & NA (NA)         \\ 
\hline
\textbf{LLC4096} & \textbf{B1-SVM} & \textbf{BP1-SVM} & \textbf{B2-SVM} & \textbf{BP2-SVM} & \textbf{A-SVM} & \textbf{P-SVM}  \\ 
\hline
Time             & 52.7 (0.8)      & 52.7 (0.8)       & 53.3 (0.8)      & 53.3 (0.8)       & 564.4 (33.4)   & 94.4 (1.3)      \\ 
\hline
TE1              & 24.5 (0.4)      & 24.4 (0.4)       & 24.7 (0.4)      & 24.6 (0.4)       & 23.9 (0.4)     & 24.3 (0.4)      \\
TE2              & NA (NA)         & 24.0 (0.5)       & NA (NA)         & 24.2 (0.5)       & NA (NA)        & 24.3 (0.4)      \\ 
\hline
Precision        & 75.1 (0.6)      & 75.5 (0.5)       & 75.0 (0.5)      & 75.3 (0.5)       & 75.8 (0.5)     & 75.3 (0.4)      \\
Recall           & 77.3 (0.6)      & 77.7 (0.5)       & 77.2 (0.5)      & 77.4 (0.4)       & 77.8 (0.5)     & 77.5 (0.4)      \\
F1               & 75.7 (0.6)      & 76.1 (0.6)       & 75.5 (0.4)      & 75.8 (0.4)       & 76.6 (0.5)     & 75.9 (0.4)      \\ 
\hline
ECE              & 14.2 (0.4)      & 14.0 (0.4)       & 14.7 (0.2)      & 14.4 (0.2)       & 13.9 (0.3)     & 14.4 (0.4)      \\ 
\hline
$k^*$       & 2               & 2                & 1               & 1                & NA (NA)        & NA (NA)         \\ 
\hline
\textbf{LLC8192} & \textbf{B1-SVM} & \textbf{BP1-SVM} & \textbf{B2-SVM} & \textbf{BP2-SVM} & \textbf{A-SVM} & \textbf{P-SVM}  \\ 
\hline
Time             & 112.4 (4.6)     & 112.4 (4.6)      & 93.7 (7.5)      & 93.8 (7.5)       & 618.6 (13.0)   & 200.1 (8.9)     \\ 
\hline
TE1              & 18.2 (0.2)      & 18.2 (0.2)       & 18.2 (0.2)      & 18.1 (0.2)       & 18.0 (0.2)     & 18.1 (0.2)      \\
TE2              & NA (NA)         & 18.1 (0.2)       & NA (NA)         & 18.0 (0.2)       & NA (NA)        & 18.1 (0.2)      \\ 
\hline
Precision        & 81.1 (0.6)      & 81.3 (0.3)       & 81.2 (0.2)      & 81.5 (0.3)       & 82.5 (0.2)     & 81.3 (0.3)      \\
Recall           & 83.7 (0.4)      & 83.8 (0.4)       & 83.7 (0.3)      & 84.0 (0.3)       & 84.3 (0.3)     & 83.9 (0.2)      \\
F1               & 81.8 (0.4)      & 82.0 (0.4)       & 81.9 (0.3)      & 82.2 (0.3)       & 82.6 (0.3)     & 81.9 (0.2)      \\ 
\hline
ECE              & 12.4 (0.3)      & 12.0 (0.4)       & 12.6 (0.4)      & 12.3 (0.4)       & 11.6 (0.2)     & 12.3 (0.4)      \\ 
\hline
$k^*$       & 2               & 2                & 1               & 1                & NA (NA)        & NA (NA)         \\
\hline
\end{tabular}}
\end{table}

Table \ref{table:table_5} reports the result for LLC encoding with ``max-L2" (the same result as ``sum-L2") pooling, which consistently gave the best performance. We denote the schemes with different codebook sizes as \textbf{LLC2048}, \textbf{LLC4096}, and \textbf{LLC8192}. Table \ref{table:table_5} suggests that LLC8192 gave the best result, surpassing the standard CNN and VQ32, but at the cost of more than a tenfold increase in computation time. LLC encoding performed better with higher-resolution images, as it could extract more locality information compared to VQ.

\subsection{SIBOW-SVM vs. Ensembles}\label{ssec:ensemble}
\blue{This section compares our proposed SIBOW-SVM method with two traditional machine learning techniques: Extreme Gradient Boosting Tree \citep[XGB,][]{Chenxgb2016} and Random Forest \citep[RF,][]{ho1995random}. These ensemble learning methods are widely used in practice for their competitive and robust performance, by aggregating predictions from multiple base classifiers \citep{tseng2023optimized,csaholczi2021automatic}. We used the same abstract image feature representation of the $\bq$-vector $\in \realset^{M}$ in the SIBOW-SVM as input for XGBoost and RF. Table \ref{table:table6}\footnote{The table shows model performance comparisons between SIBOW-SVM, CNN, Random Forest, and XGBoost. The running time is measured in minutes. The performance metrics include TE (misclassification rate), F1 (macro average), and ECE (expected calibration error) multiplied by 100 for both mean and standard derivation (displayed in parentheses). $p$-values were calculated using the Wilcoxon paired signed-rank test.} presents their estimation accuracy and computation time. }

\blue{To assess the statistical significance of performance differences among the various methods, we employed a Wilcoxon paired signed-rank test. This nonparametric statistical test is particularly well-suited for comparing paired observations. Table 6 reports P-values for comparing each method against two SIBOW-SVM methods: using VQ64 SIBOW-SVM as a benchmark in the top half of the table, and using LLC8192 SIBOW-SVM as a benchmark in the bottom half. 
Results indicate that the VQ64 SIBOW-SVM significantly outperforms traditional methods across all performance measures, including classification error, F-1 score, and ECE, with $p$-values below $2\times 10^{-16}$. And LLC8192 SIBOW-SVM also outperforms traditional methods across all measures with $p$-values below $0.05$. This analysis suggests that SIBOW-SVMs offer competitive performance both on multi-class classification and class probability estimation, with the VQ64 SIBOW-SVM exhibiting superior performance among all compared models.}  

\begin{table}[!htbp]
\caption{Model performance comparisons}
\label{table:table6}
\centering
\setlength{\tabcolsep}{3pt}
\scalebox{0.75}{
\begin{tabular}{c|c|c|c|c} 
\hline
\textbf{Metrics} & \textbf{VQ64}    & \textbf{CNN1} & \textbf{RF} & \textbf{XGB}  \\ 
\hline
Time             & 9.5 (0.7)        & 25.4 (1.8)    & 1.5 (0.1)   & 160.5 (4.5)       \\
$p$-value          & NA               & $<2e\!-\!16$       & $<2e\!-\!16$      & $<2e\!-\!16$            \\ 
\hline
TE               & 13.9 (0.2)       & 22.7 (0.4)    & ~26.1 (0.2) & 25.4 (0.2)        \\
$p$-value          & NA               & $<2e\!-\!16$        & $<2e\!-\!16$      & $<2e\!-\!16$            \\ 
\hline
F1               & 87.2 (0.2)       & 76.7 (0.4)    & 71.1 (0.4)  & 73.1 (0.5)        \\
$p$-value          & NA               & $<2e\!-\!16$        & $<2e\!-\!16$      & $<2e\!-\!16$            \\ 
\hline
ECE              & 9.0 (0.2)        & 20.2 (0.3)    & 24.7 (0.3)  & 22.1 (0.2)        \\
$p$-value          & NA               & $<2e\!-\!16$        & $<2e\!-\!16$      & $<2e\!-\!16$            \\ 
\hline
\textbf{Metrics} & \textbf{LLC8192} & \textbf{CNN1} & \textbf{RF} & \textbf{XGB}  \\ 
\hline
Time             & 93.8 (7.5)       & 25.4 (1.8)    & 13.5 (0.2)  & 478.3 (5.2)       \\
$p$-value          & NA               & $2.94e\!-\!3$     & $1.37e\!-\!4$   & $<2e\!-\!16$            \\ 
\hline
TE               & 18.0 (0.2)       & 22.7 (0.4)    & 29.2 (0.4)  & 25.7 (0.3)        \\
$p$-value          & NA               & $1.21e\!-\!2$       & $<2e\!-\!16$      & $<2e\!-\!16$            \\ 
\hline
F1               & 82.2 (0.3)       & 76.7 (0.4)    & 68.4 (0.3)  & 71.5 (0.4)        \\
$p$-value          & NA               & $1.05e\!-\!3$       & $<2e\!-\!16$      & $<2e\!-\!16$            \\ 
\hline
ECE              & 12.3 (0.4)       & 20.2 (0.3)    & 29.1 (0.3)  & 25.4 (0.2)        \\
$p$-value          & NA               & $<2e\!-\!16$        & $<2e\!-\!16$      & $<2e\!-\!16$            \\
\hline
\end{tabular}}
\end{table}

\subsection{Discussion}\label{ssec:4.3}
This article introduces the SIBOW-SVM model, which incorporates two encoding schemes, VQ and LLC, for classifying multi-dimensional images. These novel methods demonstrate significant improvements over standard CNN architectures in the context of brain tumor MRI image classification and probability estimation, particularly when combined with appropriate encoding and pooling methods for final image feature representation. 

\begin{figure}[!htbp]
\centering
     \includegraphics[width=0.85\textwidth, height=0.7\textheight, keepaspectratio]{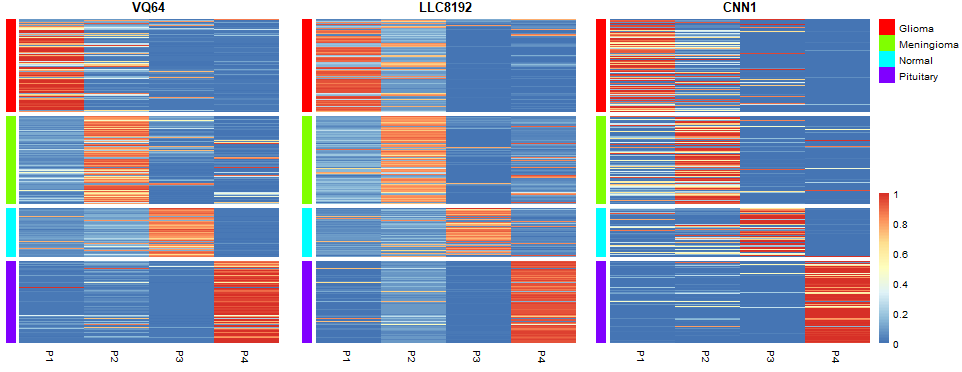}
\caption{This figure plots the average multiclass probability estimation heat map for VQ64 SIBOW-BPSVM, LLC8192 SIBOW-BPSVM, and CNN1, based on 10 simulations.}
\label{fig:figure_3}
\end{figure}

\begin{figure}[!htbp]
\centering
     \includegraphics[width=0.85\textwidth, height=0.7\textheight, keepaspectratio]{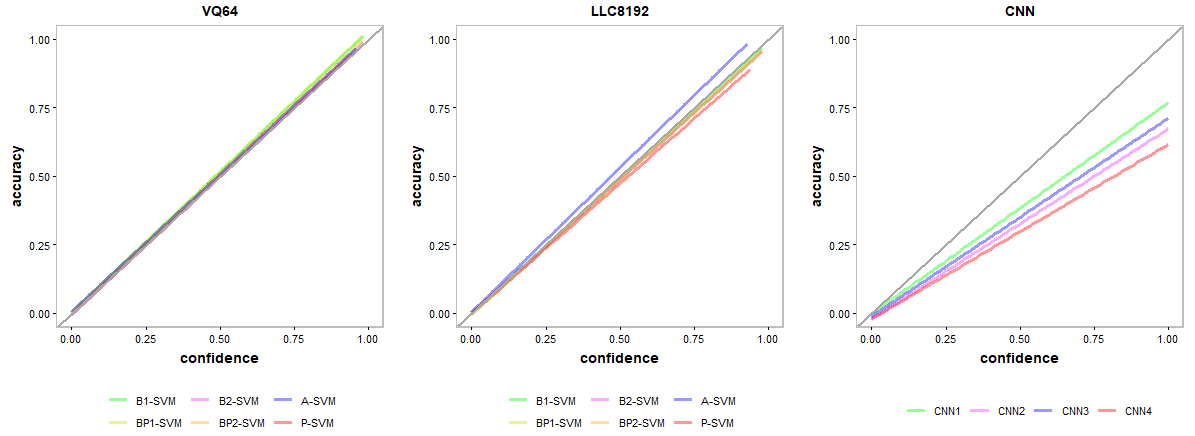}
\caption{This figure plots the reliability diagrams for VQ64, LLC8192 and 4 CNN architectures. The reliability curves have been smoothed using GAM. Different colors represent different methods. A close proximity with the diagonal line indicates accurate model calibration.}
\label{fig:figure_4}
\end{figure}

We conducted a comprehensive comparison between the two top-performing SIBOW-SVM model schemes, VQ64 and LLC8192, and the standard CNN architecture to evaluate multiclass probability estimation and classification. Fig. \ref{fig:figure_3} presents a heat map of probability estimates for the four tumor classes, illustrating that our new methods yield superior class probability estimation to CNNs. Figure \ref{fig:figure_4} displays a reliability diagram showing the quality of class probability estimation. The SIBOW-SVM consistently exhibits better model calibration by adhering more closely to the ${45}\degree$ diagonal line, outperforming the four CNNs. Notably, the VQ64 scheme achieves nearly perfect calibration across all wSVM methods. 

\begin{figure}[!htbp]
\centering
     \includegraphics[width=0.85\textwidth, height=0.7\textheight, keepaspectratio]{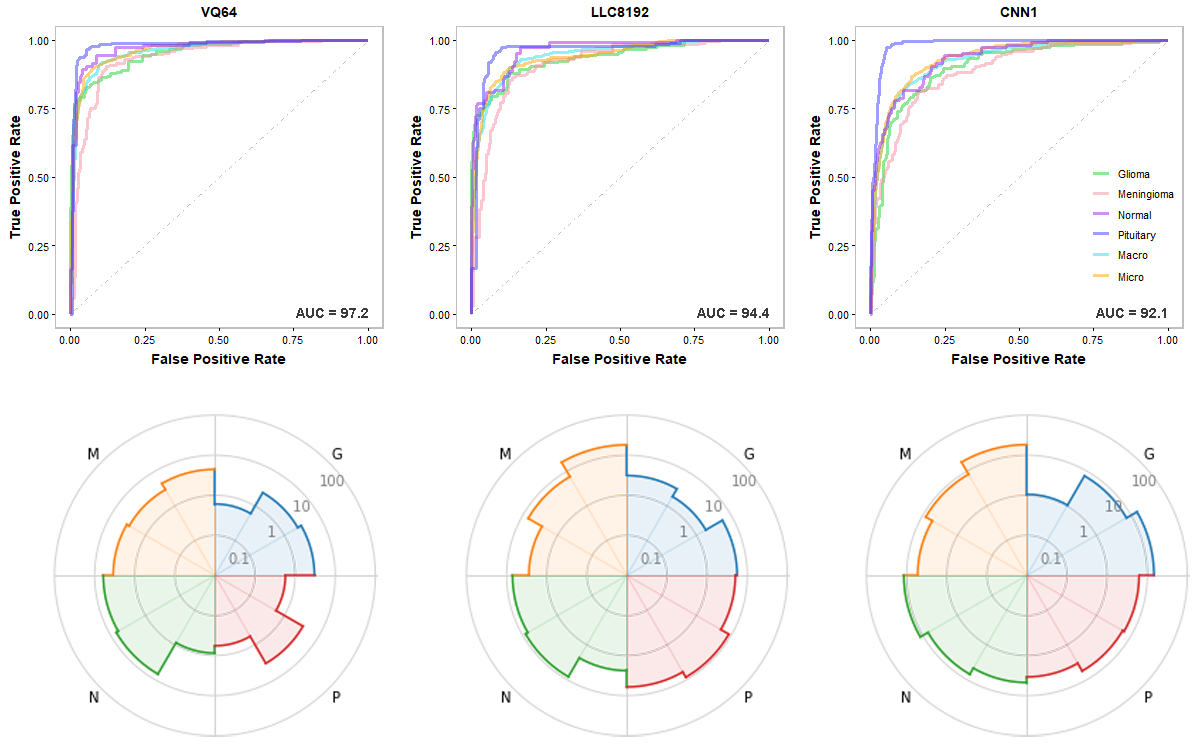}
\caption{This figure plots the multiclass ROC and confusion stars for VQ64 SIBOW-BPSVM, LLC8192 SIBOW-BPSVM, and CNN1. A higher multiclass AUC value indicates a better classifier. A smaller area of the confusion star suggests a lower error in the multiclass confusion matrix.}
\label{fig:figure_5}
\end{figure}

Fig. \ref{fig:figure_5} presents multiclass receiver operating characteristic (ROC) curves and the confusion star as a visual representation of the confusion matrix \citep{Luque9658486}. Our methods excel in multiclass classification, achieving higher multiclass AUC scores defined by \citet{hand_till_2001} and smaller polar lengths as an error measure for the confusion matrix. 

\begin{figure}[!htbp]
\centering
     \includegraphics[clip, trim=0cm 0cm 0cm 0cm, width=0.85\textwidth, height=0.515\textheight, keepaspectratio]{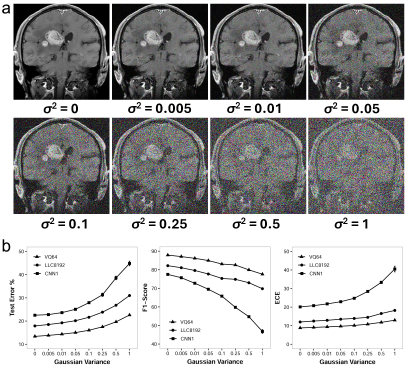}
\caption{\blue{This figure illustrates the robustness of three image classification methods: VQ64 SIBOW-BPSVM, LLC8192 SIBOW-BPSVM, and CNN1, when random Gaussian noise with varying
variance $\sigma^2$ is added to the original MRI images, as illustrates in (a). Figure (b) illustrates the robustness measure for TE (misclassification rate) shown as a percentage, F1 (macro average) and ECE (expected calibration error) multiplied by 100. Different point shapes represent different methods. A smaller change in slope across variances indicates greater robustness to image perturbations.}}
\label{fig:figure_6}
\end{figure}

\blue{In addition to reporting classification accuracy, we have expanded our evaluation beyond classification accuracy to include a new set of numerical experiments. These experiments assess and compare the data robustness of various methods when faced with contaminated or noisy data due to measurement errors. Specifically, we perturbed the original image set with 10 random splits described in Section \ref{ssec:4.1} and \ref{ssec:4.2} by introducing Gaussian noise at various levels and compared all methods under these conditions. Figure \ref{fig:figure_6} illustrates the performance of all models in terms of classification error and probability estimation when the input images are subjected to Gaussian noise with $\mu =0$ and $\sigma^2 \in \{0, 0.005, 0.01, 0.05, 0.1, 0.25, 0.5, 1\}$. Our results show that the proposed SIBOW-SVM methods generally demonstrate more robust performance against image noise compared to standard CNNs.} Notably, VQ64 SIBOW-SVM consistently outperforms all other methods across different noise levels. These findings highlight the potential of VQ64 for challenging image classification tasks, particularly in scenarios where high levels of noise are present in the data.

In the literature, it has been demonstrated that soft-VQ encoding methods such as LLC offer better performance in image classification tasks \citep{Wang5540018, agustsson2017}, due to their ability to capture more locality information. However, LLC typically requires a large codebook with high-dimensional
final image features, which can slow down computations. Our findings suggest that, VQ encoding, when paired with appropriate encoding and pooling methods, can achieve better performance within the wSVMs framework for probability estimation while maintaining reduced computational complexity. Furthermore, the downstream wSVMs framework can be significantly accelerated through parallel computing, leveraging its divide-and-conquer nature, making it a competitive alternative to CNNs with NVIDIA CUDA GPU acceleration.

\section{Concluding Remarks}\label{sec:cond}
The proposed SIBOW-SVM framework shows promise as a method for image classification and confidence measurement of predictions. The wSVMs with baseline learning exhibit overall high accuracy and computational efficiency, making it a favorable approach for multiclass probability estimation in image classifications. The SIBOW-SVM, when coupled with appropriate encoding and pooling methods, consistently provides robust and improved class probability estimation and image classification compared to standard CNNs.

For further work, we plan to enhance the SIBOW-SVM by incorporating sparse learning capabilities. In practice, images with high noise and low resolution are common, leading to the extraction of noisy features by the SIFT algorithm. The current SIBOW-SVM only works effectively with dense features, and its performance may significantly deteriorate when dealing with sparse image features. To address this challenge, we may introduce $\ell^1$-norm regularization to the SIBOW-wSVMs, enabling automatic feature selection and image classification in the presence of numerous noise features. 


\bibliographystyle{plainnat}
\bibliography{reflist}





\end{document}